%% file: RRLyr_Kepler_MNRAS_refereed.tex
\documentclass[useAMS,usenatbib]{mn2e}
\usepackage{graphicx}

\title[{\it Kepler} photometry of the prototypical Blazhko star RR~Lyr]{{\it Kepler} photometry of the prototypical Blazhko star RR~Lyr:\\
\hspace{3.5cm} An old friend seen in a new light
}
\author[K. Kolenberg, et al.]
{K. Kolenberg$^{1}$\thanks{E-mail:
katrien.kolenberg@univie.ac.at}
S. Bryson$^{2}$,
R. Szab\'o$^{3}$,
D. W. Kurtz$^{4}$,
R. Smolec$^{1}$,
J.~M. Nemec$^{5}$,
\and
E. Guggenberger$^{1}$,
P. Moskalik$^{6}$,
J.~M. Benk\H{o}$^{3}$,
M. Chadid$^{7}$,
Y.-B. Jeon$^{8}$,
L.~L. Kiss$^{3,9}$,
\and
G. Kopacki$^{10}$,
J. Nuspl$^{3}$,
M. Still$^{2}$,
J. Christensen-Dalsgaard$^{11}$, H. Kjeldsen$^{11}$, 
\and 
W. J. Borucki$^{2}$, 
D. A. Caldwell$^{12}$, J. M. Jenkins$^{12}$, D. Koch$^{2}$
\\
$^{1}$Institut f\"ur Astronomie, University of Vienna, T\"urkenschanzstrasse 17, A-1180 Vienna, Austria\\
$^{2}$NASA Ames Research Center, MS 244-30, Moffet Field, CA 94035, USA\\
$^{3}$Konkoly Observatory of the Hungarian Academy of Sciences, Quarter Thege Mikl\'os \'ut 15-17, H-1121 Budapest, Hungary\\
$^{4}$Jeremiah Horrocks Institute of Astrophysics, University of Central Lancashire, Preston PR1 2HE, UK\\
$^{5}$Department of Physics \& Astronomy, Camosun College, Victoria, British Columbia, V8P 5J2, Canada\\
$^{6}$Copernicus Astronomical Center, ul. Bartycka 18, 00-716 Warsaw, Poland\\
$^{7}$Observatoire de la C\^ote dÕAzur, Universit\'e Nice Sophia-Antipolis, UMR 6525, Parc Valrose, 06108 Nice Cedex 02, France\\
$^{8}$Korea Astronomy and Space Science Institute, Daejeon, 305-348, Korea\\
$^{9}$Sydney Insitute for Astronomy, School of Physics, University of Sydney, NSW 2006, Australia\\
$^{10}$Instytut Astronomiczny Uniwersytetu Wrocl{}awskiego, Kopernika 11, 
       51-622 Wrocl{}aw, Poland\\
  $^{11}$Department of Physics and Astronomy, Aarhus University, DK-8000 Aarhus C, Denmark\\
$^{12}$SETI Institute, Mountain View, CA 94043, USA\\}
\begin{document}

\date{Accepted 2010, ? Received 2010, August}

\maketitle

\begin{abstract}
We present our analysis of the long cadence {\it Kepler} data for the well-studied Blazhko star RR~Lyr, gathered during the first two quarters of the satellite's observations and covering a total of 127\,d. 
Besides being of great importance for our understanding of RR Lyrae stars in general, these RR~Lyr data can be regarded as a case study for observations of bright stars with {\it Kepler}. 
{\it Kepler} can perform high-precision photometry on targets like RR Lyr, as the saturated flux is conserved to a very high degree.  
The {\it Kepler} data on RR~Lyr are revolutionary in several respects.  
Even with long-cadence sampling (one measurement per 29.4\,min), the unprecedented precision ( $<$ mmag) of the {\it Kepler} photometry allows the study of the star's extreme light curve variations in detail.  The multiplet structures at the main frequency and its harmonics, typical for Blazhko stars, are clearly detected up to the quintuplets.  For the first time, photometric data of RR~Lyr reveal the presence of half-integer frequencies, 
linked to a period doubling effect.  This phenomenon may be connected to the still unexplained Blazhko modulation. 
Moreover, with three observed Blazhko cycles at our disposal, we observe that there is no exact repetition in the light curve changes from one modulation cycle to the next for RR~Lyr.  This may be due to additional periodicities in the star, or to transient or quasi-periodic changes.
\end{abstract}


\section{Introduction}
The pulsating star RR~Lyr has been studied for over a century now (Szeidl \& Koll\'ath 2000).  In 1948 this field star became the eponym of a class of variables previously called ``cluster variables" (see Smith 1995).
RR\,Lyrae stars are low-mass, horizontal branch stars (post-red giant stage) that are 
burning helium in their core. They play
a key role in astrophysics for their use as
standard candles, as examples of stars in an advanced evolutionary stage,
and because of their ``simple"
radial pulsations.  RR\,Lyrae stars have typical periods 
of $\sim$0.2\,d to $\sim$1\,d, amplitudes in the optical of 0.3\,mag up to 2\,mag, and spectral types 
of A2 to F6. They pulsate in the radial fundamental mode (RRab 
stars), the radial first overtone (RRc stars) and, in some cases, in both modes 
simultaneously (RRd stars). A few RR~Lyrae stars are suspected to be pulsating in higher-order radial overtone modes.

A large fraction of RR Lyrae stars shows the Blazhko effect (Blazhko 1907), a periodic amplitude and/or phase
modulation with a period of typically tens to hundreds of times the pulsation
period. 
The origin of the Blazhko effect remains a matter of controversy.  
Rotational modulation of the distorted radial mode invoked by a strong dipole-like magnetic field (Shibahashi 2000) appears to have been ruled out  as a cause(Chadid et al. 2004; Kolenberg \& Bagnulo 2009).  Several models proposing a  resonance between the main radial mode and another mode have been proposed -- with another radial mode (higher overtone: Borkowski 1980, Moskalik 1986), or with a nonradial mode of low spherical degree $\ell$ (Van Hoolst et al. 1998; Dziembowski \& Mizerski 2004) -- but their validity remains to be proven. 
Stothers (2006, 2010) proposes a scenario that attributes the Blazhko variation to variable turbulent convection caused by, e.g., transient magnetic fields in the star (that would be hard to detect).
Thus far, all models presently proposed for the Blazhko effect have shortcomings in explaining the variety of features shown by modulated stars (Kov\'acs 2009). 

The RRab star RR~Lyr is the brightest star, eponym and prototype of its class, and it shows very strong Blazhko modulation. 
It has been extensively studied over the course of the past century, through photometric (e.g., Szeidl \& Koll\'ath 2000), spectroscopic (e.g., Preston et al. 1965), and spectropolarimetric  (e.g., Chadid et al. 2004) data. 
  However, several aspects of its pulsation remain poorly understood,
  and to accurately model the star, we have to take into account complex physics that we are only beginning to uncover.  
    
  By a fortunate coincidence, RR~Lyr ($\alpha(J2000)$: 19$^{\rm h}$ 25$^{\rm m}$ 28$^{\rm
  s}$ and $\delta(J2000)$: +42$^{\circ}$ 47$^{\prime}$ 04$^{\prime\prime}$) lies in the field of the {\it Kepler} space telescope ({\it Kepler} ID: KIC7198959), though its brightness is well above the saturation limit for the {\it Kepler} CCDs.
  The unprecedented quality of the {\it Kepler} data of the star leads us to several new findings that spur further modelling.
  To explore theoretical models of RR Lyr, it is necessary to know the physical parameters of the star with the highest accuracy possible.
For this reason, a self-consistent atmospheric and abundance analysis from high-resolution, high signal-to-noise spectra of the star, with the aim of parameter determination, was recently published by Kolenberg et al. (2010b).

  According to the General Catalogue of Variable Stars (Samus 2004) 
its $V$ brightness changes within the range 7.06 -- 8.12 mag and its period is $P$ = 0.56686776 d 
or about 13 h 36 min, corresponding to a frequency of $f_0 = 1.7640799$ d$^{\rm -1}$.
Kolenberg et al. (2006) found a (mean) frequency of $f_0 = 1.764170 \pm 0.000005$ d$^{\rm -1}$ from data gathered in 2003 and 2004.

In Section\,2 we present the new observations of RR~Lyr obtained with the {\it Kepler} satellite and describe the challenges of the data reduction. Section\,3 presents the data analysis and its results.  In Section\,4 we discuss our results in the framework of the current understanding of RR Lyrae pulsations and the Blazhko effect. Finally, some concluding remarks and the outlook for future investigations are given in Section\,5.

\section{Observations}\label{obs}

The {\it Kepler} space telescope (Borucki et al. 2010) was designed to detect transits of Earth-like planets around Sun-like stars.  Technical details on the mission can be found in Koch et al. (2010) and Jenkins et al. (2010a,b). In order to exploit the potential of the {\it Kepler} asteroseismic data (Gilliland et al. 2010), the Kepler Asteroseismic Science Consortium (KASC) was set up, grouping more than 350 scientists from all over the world.  KASC working group 13 is dedicated to the study of the {\it Kepler} RR Lyrae stars.  At the time of writing, over 30 RR Lyrae stars have been identified in the {\it Kepler} field and are now observed by the satellite. About half of the {\it Kepler} RR Lyrae targets have turned out to be modulated (Kolenberg et al. 2010a; Benk\H{o} et al. 2010). With {\it Kepler}'s unprecedented precision, we will find additional clues to solving the Blazhko riddle. 

\begin{figure*}
\includegraphics[width=18cm]{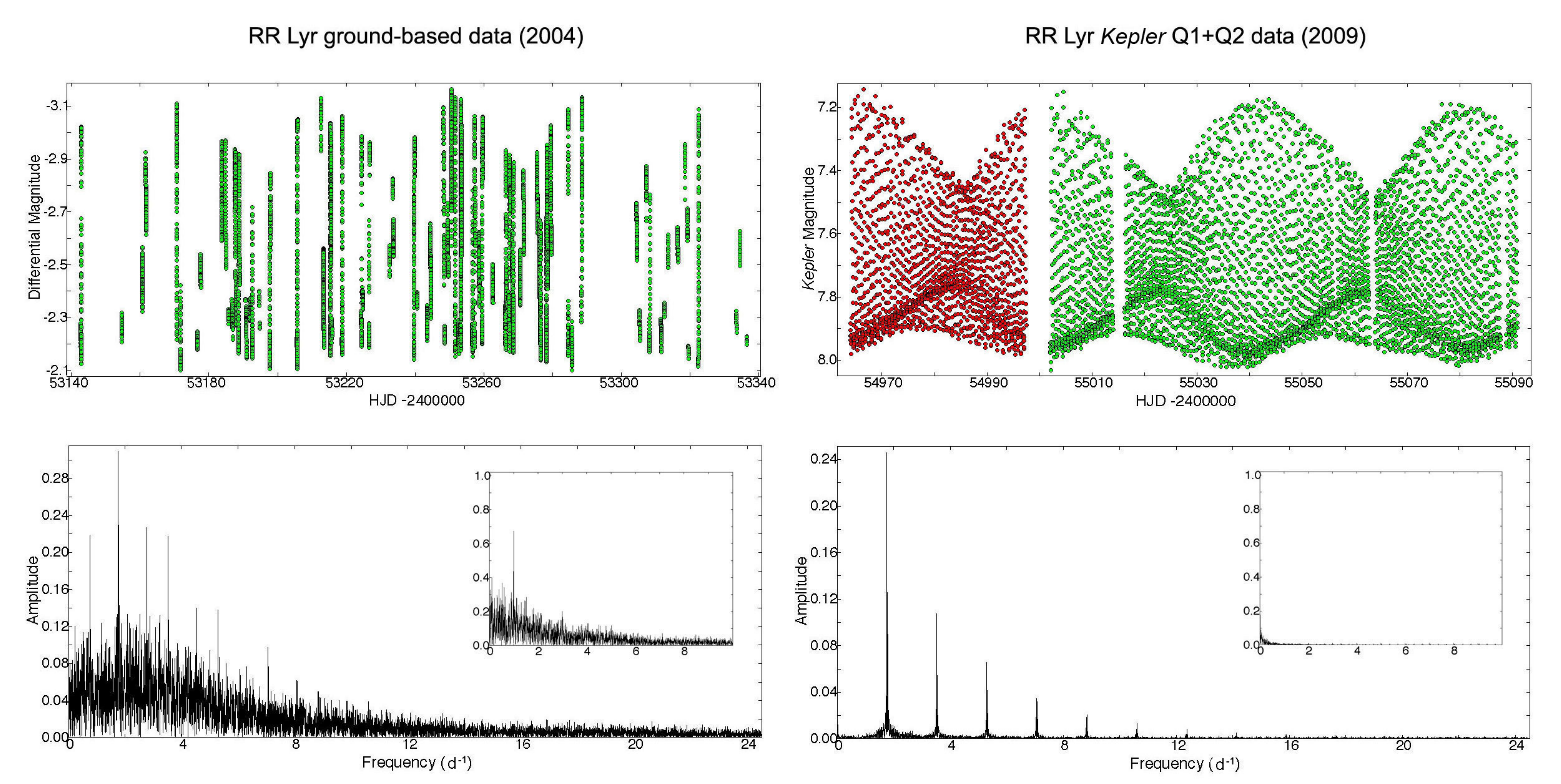}
 \caption{Comparison of the ground-based RR~Lyr data, gathered from six different observatories, published by Kolenberg et al. (2006) and the {\it Kepler} Q1 and Q2 data of the star transformed to the magnitude scale (top panels).  Individual cycles are shown in Fig.\,\ref{rrLyraeCorrection}. Bottom panels: Fourier transform of the data; the insert shows the window function. \label{RR_Kp}}
\end{figure*}

The {\it Kepler} Mission offers two options for observations: either long cadence (29.4 minutes) or short cadence (1 minute). 
The spacecraft ``rolls" every three months to allow for continuous illumination of {\it Kepler}'s solar arrays.  RR~Lyr was observed in long-cadence during the first roll of the 
{\it Kepler} survey phase between HJD\,2454964.0109 and HJD\,2454997.4812 
(2009 May 12 -- June 14; Q1 data, 33.5\,d, 1628 useful data points), and during the second roll between HJD\,55002.01748 and  HJD\,55090.96492 (2009 June 19 -- September 16; Q2 data 88.8\,d, 4097 useful data points). The total time span of the data set is nearly 127\,d, more than 3 complete Blazhko cycles, as can be seen in Fig.\,\ref{RR_Kp}. There are a few small gaps in the data set, due to unplanned safe mode, loss of fine pointing events as well as regular data downlink periods. Note that except for around those gaps,  nearly every single {\it Kepler} measurement is of excellent quality.
Long-cadence sampling is just sufficient to obtain a good coverage of the light curve, but rapid changes and ``glitches" in the light curve are missed by this sampling, and can be investigated with forthcoming short-cadence {\it Kepler} data. As a consequence of the sampling rate, the Nyquist frequency lies at 
24.5\,d$^{-1}$. 

Fig.\,\ref{RR_Kp} shows the quality of the {\it Kepler} data on RR~Lyr compared to one of the most precise published data sets on the star obtained from ground-based observatories (Kolenberg et al. 2006). Despite the effort of organizing a multi-site campaign and combining all the standardized observations, the latter data set had a point-to-point scatter of over 0.005 mag at best, and was characterized by nightly and weather-related gaps, typical for Earth-based observations.
As a consequence, the Fourier spectra of such data sets obtained from the ground have a noise level 10--50 times higher than {\it Kepler} Fourier spectra and are subject to aliasing. 

\subsection{Early {\it Kepler} photometry of RR~Lyr}\label{phot}

\begin{figure}
\center
\includegraphics[height=90mm]{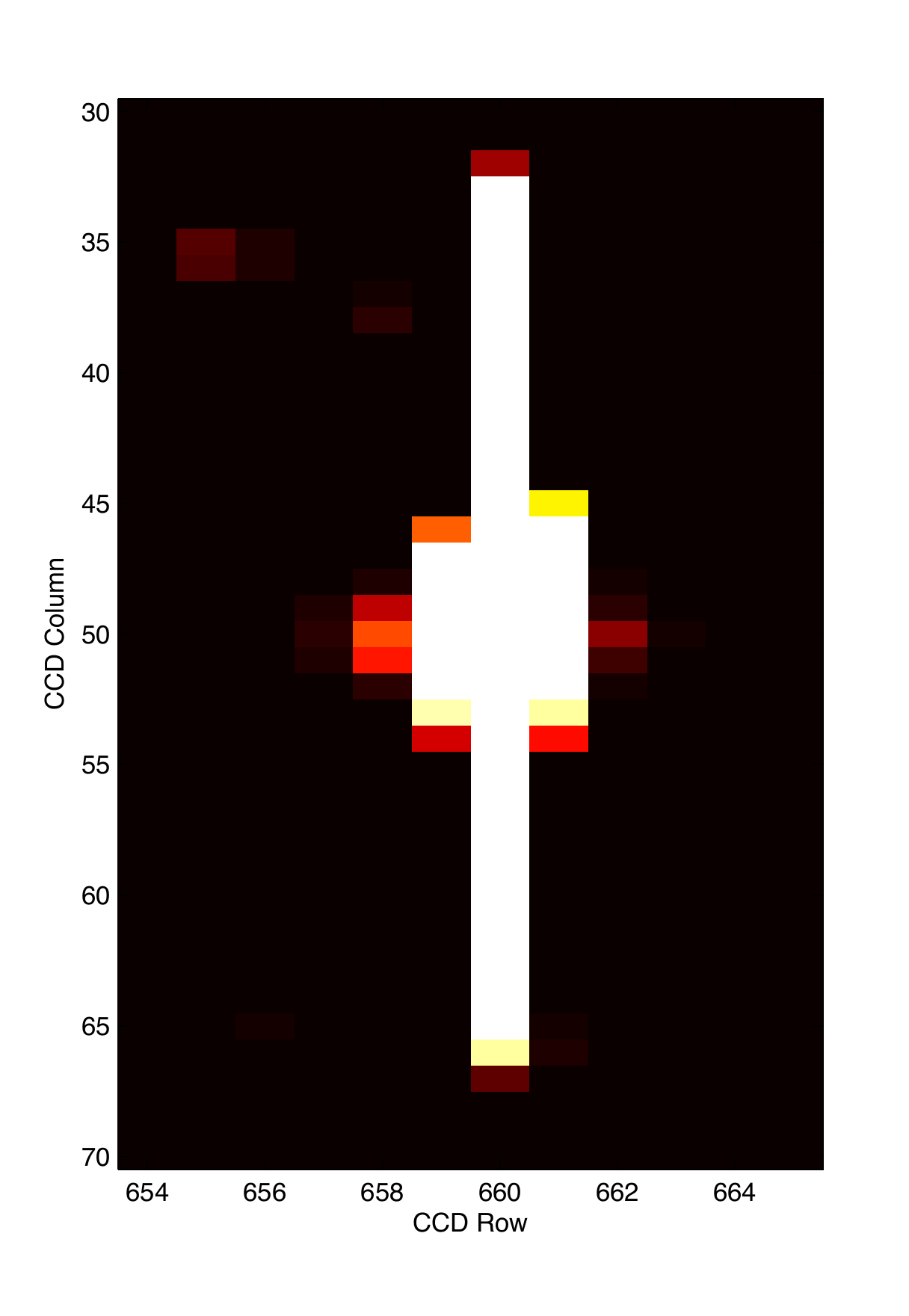}
\caption{RR~Lyr's downlinked pixels during quarter 2 (Q2), showing the saturation spill in the three central columns.   The first cadence in Q2 is shown, when RR~Lyr was near a minimum.  For brighter cadences the central column saturation exceeded these pixels, but the two adjacent columns remained completely captured.  Note the differing row and column scales.}
\label{rrLyraePixels}
\end{figure}

{\it Kepler}'s CCDs saturate between {\it Kepler} magnitudes 11 and 12, depending on which CCD a target falls, as well as the target location on that CCD.  This saturation spills up and down the CCD column, and saturated flux is conserved to a very high degree (see Fig.~\ref{rrLyraePixels}).  Therefore, {\it Kepler} can perform high-precision photometry on saturated targets like RR~Lyr.  Due to bandwidth constraints, {\it Kepler} downlinks pixels selected for each target star (Haas et al. 2010).  These pixels are selected algorithmically to maximize the signal-to-noise ratio for the target (Bryson et al. 2010), based on stellar properties in the {\it Kepler} Input Catalog (KIC) and various models of sky and spacecraft characteristics, including saturation behavior.  In the case of RR~Lyr, the {\it Kepler} magnitude in the KIC did not reflect maximum brightness.  This, combined with the early and approximate state of the saturation model at the time of Q1 and Q2 observations, resulted in the size of RR~Lyr's saturation being underestimated.  Therefore, not enough pixels were allocated to RR~Lyr to fully capture all flux at its brightest maxima:  the central saturation column exceeded the assigned aperture.  In Q1 the consequent loss of flux was relatively small but in Q2 the loss of flux at maxima was significant.

Fortunately, columns adjacent to the central saturation column also saturated, though to a lesser extent, and these columns were fully captured in all of Q1 and Q2.  This allows us to use those measurements where the central column is fully captured (away from maxima) to determine the ratio of the central columnÕs flux to the flux in the adjacent columns, and use this ratio as a predictor of the flux in the central column when the central column is not fully captured.  This ratio, $r_0$, is sensitive to various known pointing and focus systematics, so we determine $r_0$ for every measurement, and detrend $r_0$ using a robust piecewise-polynomial fit to those measurements where the central column's flux is fully captured.  The resulting polynomial trend $r_t$ was consistent with known motion and focus systematics.  We define a correction $c=r_t/r_0$  which multiplies the central column's flux for those data points in which some of the 
flux falls outside RR~LyrÕs pixels, as determined by the detrended ratio $r_d$ falling below a cutoff value. In Q1 only minor corrections on the order of 10 per cent were required for the central column flux, and these were at the beginning and end of Q1 (from MJD 54964.25 to 54972.76 and 54995.47 to 54997.17).  In Q2 corrections to the central column's flux were required for every maximum and ranged from 30 to 60 per cent.  The uncertainty in the correction is estimated, from the variance of $r_d$, to be 0.45 per cent in Q1 and 0.36 per cent in Q2 (neglecting six measurements with large outliers in Q2). 

Once the central column flux is corrected, it is summed with the other columns in the aperture to create the estimated flux correction for each measurement.  This flux correction is applied to the output of the {\it Kepler} Science Operations Center pipeline, which corrects light curves by removing cosmic ray hits, background and correlations with known systematics (Jenkins et al. 2010a,b). 

\begin{figure}
\includegraphics[height=70mm]{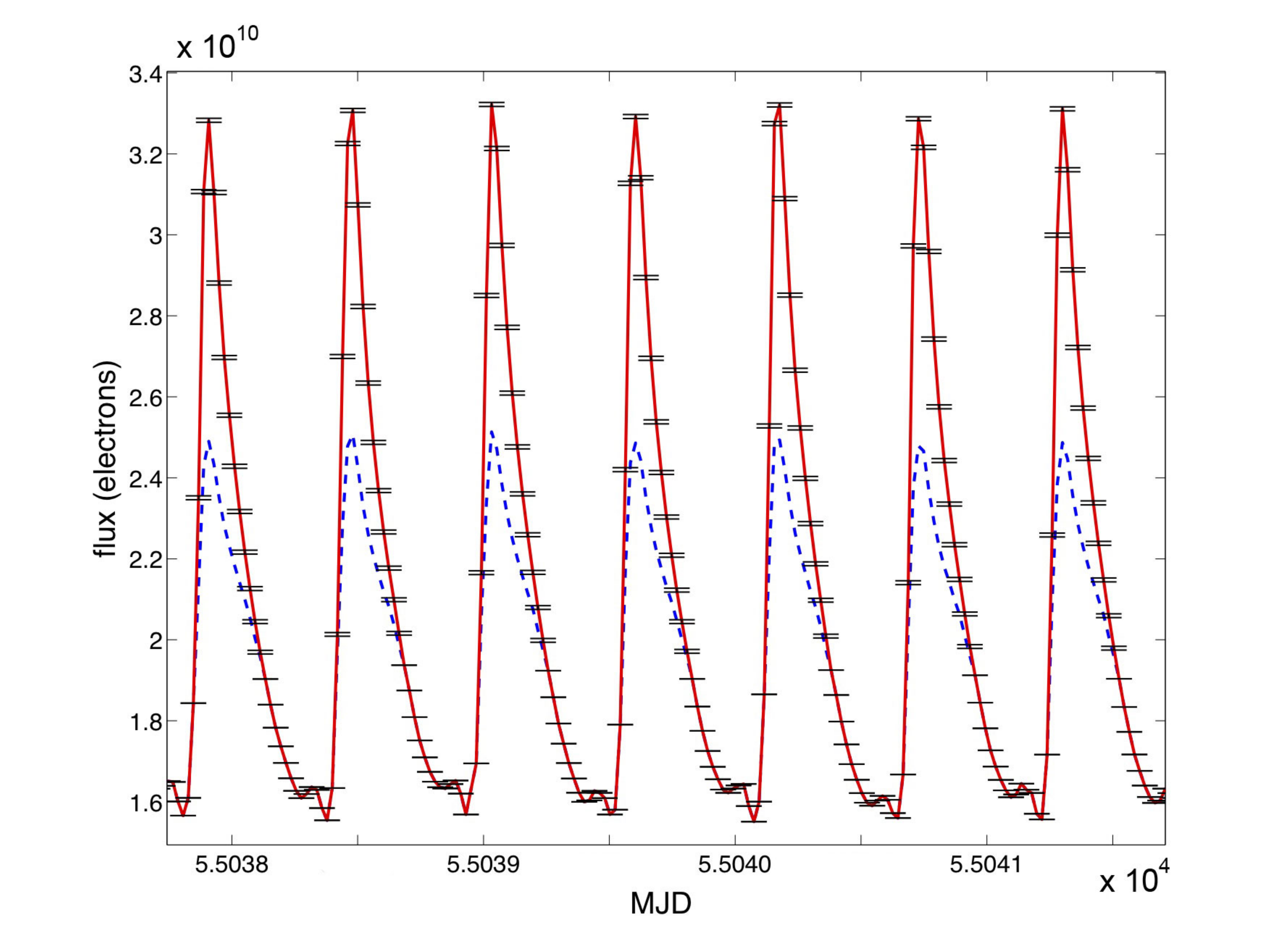}
\caption{Comparison between the captured flux from RR~Lyr (dashed line) and the corrected flux (full line) during Q2, with $\pm1$ $\sigma$ error bars. }
\label{rrLyraeCorrection}
\end{figure}

An example of the original and corrected light curves in Q2 is shown in Fig.~\ref{rrLyraeCorrection}.  In Q1 the final flux corrections were on the order of at most 5 per cent.  In Q2 the flux corrections range from 15 to 35 per cent. 
The uncertainty in the total corrected flux is about 0.25 per cent in Q1 and Q2 for measurements in which the correction was applied, compared with the flux uncertainty of $8\times10^{-6}$ in measurements where the correction was not applied because the flux was completely captured.  

\subsection[]{RR~Lyr's {\it Kepler} Magnitude}\label{phot}

{\it Kepler} processing does not provide calibrated {\it Kepler} magnitudes $Kp$.  We produced a rough estimate of RR Lyr's magnitude in Q1 and Q2 using 5 nearby quiet (95 per cent variation under 1 part per thousand) stars of known $Kp$.  Because this method did not account for colour or uncertainties in the comparison star's magnitudes and measured flux, the $Kp$ magnitude values here are only approximate.  We emphasize, however, that the relative precision within
a given quarter is unprecedentedly high, the approximate nature and ambiguity is only in tying the RR~Lyr light curve to the absolute {\it Kepler} photometry.

As can be seen in Fig.\,\ref{RR_Kp}'s upper right panel, RR~Lyr's total light curve amplitude varies significantly over the Blazhko cycle.  The total variation (from lowest minimum to highest maximum) spans 0.86 mag in $Kp$, 
with 0.84 mag for the largest amplitude light curves,
and 0.47 mag for the smallest amplitude light curves.
Hence, the total reduction of the light curve amplitude due to the Blazhko effect is about 44 per cent as determined from our {\it Kepler} data of RR~Lyr so far.  These values will be refined with the short cadence data to come.

The GCVS lists a 7.06 -- 8.12 (1.06 mag; see also total amplitude in Fig.\,\ref{RR_Kp}'s left upper panel) variation in $V$ magnitude.  Hence, with the wide (white-light) passband of {\it Kepler} we observe an amplitude reduction with respect to the $V$ magnitudes by a factor of 1.23.  The same factor is found when comparing the amplitude $A_1$ of $f_0$ (the first Fourier component) in the $V$ data published by Kolenberg et al. (2006) to the one obtained from our data (Table\,\ref{RR_freq}).  




The possible deviations from the ``real" {\it Kepler}
magnitudes 
introduced by the corrections and their associated uncertainties have a
negligible effect on the frequency
values.  For Q2, the quarter covering more than two Blazhko cycles,
and for which the largest corrections
were needed, the frequency values derived from the non-corrected and
corrected data are identical to within the
error bars.   For the Q1 data this is also the case but the comparison is
less convincing since they cover less than one Blazhko cycle.

\section{Data analysis and results }\label{analysis}

A detailed visual inspection of the {\it Kepler} data of RR~Lyr has already revealed changes between consecutive Blazhko cycles, as well as alternating higher and lower light curve maxima at certain phases in the Blazhko cycle (see fig.\,4 in Kolenberg et al. 2010a).  Because of this, we can expect a more complex frequency spectrum than the ``classical" {\it ``main frequency + harmonics + equidistant multiplet structures"} predicted for a monoperiodic star undergoing amplitude and phase modulation (Szeidl \& Jurcsik 2009; Benk\H{o} et al. 2009).  Moreover, the frequency solution we obtain partly depends on the coverage, as long-term modulation, secular trends, or transient phenomena cannot be fully captured by the Fourier techniques we adopt.
The Fourier analyses presented here were performed with \textsc{Period04} (Lenz \& Breger 2005).

Without adjustments, there is a small shift between the Q1 and Q2 data even after the calibrations described in Section\,\ref{obs}. To create the best match between the two data sets (as shown in Fig.\,\ref{RR_Kp}), we fitted the complete frequency solution (described in the next sections) to each set separately and subsequently shifted by the zero point of the obtained fit. 
We used the combined data set (Q1 and Q2) to determine frequencies, as a longer time base and better coverage of the Blazhko cycle yields more accurate (mean) frequency values. Though we do not have sufficient information to reliably combine the Q1 and Q2 data regarding their {\it amplitudes}, the obtained frequency values are not significantly influenced by the small zero point shifts.  Subsequently, we applied the harmonic fit with the frequencies derived from Q1 and Q2 to the longest data set (Q2). The results are summarized in Table~\ref{RR_freq}, and discussed below.  Optimum values for the frequencies, amplitudes and phases were obtained by minimizing the residuals of the fit.  Errors were determined through Monte Carlo simulations. 

\begin{figure*}
\includegraphics[width=18cm]{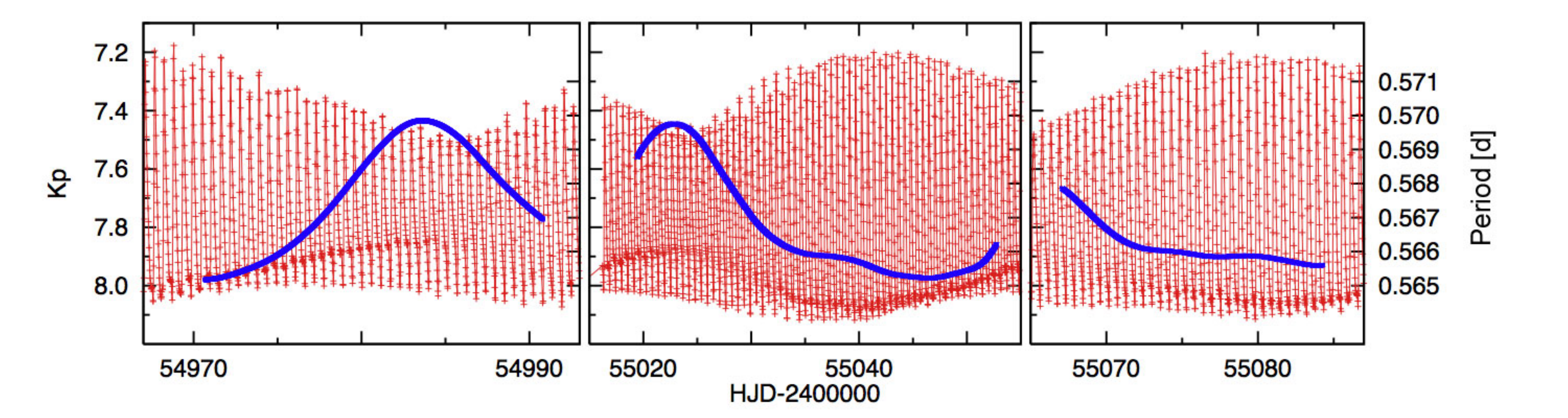}
 \caption{Relationship between the period change and amplitude modulation over the Blazhko cycle for RR~Lyr. The crosses show the {\it Kepler} data for the three monitored Blazhko cycles, the connecting line our multi-frequency fit to the data. The thick line shows an instantaneous period determined using the analytical signal method (units to the right side of the panel).  The period
is longest just before the lowest amplitude phases and shortest just after
the highest amplitude phases. Note the small changes in the period variation between consecutive Blazhko cycles, and the alternating high and low maxima at certain Blazhko phases. \label{peramp}}
\end{figure*}

\input{RR_tab1.tex}\label{freqtab}

\subsection{Typical Blazhko patterns}

RR~Lyr's high-amplitude (nonlinear) pulsation gives rise to a non-sinusoidal light curve, described in Fourier space by contributions from the main frequency $f_0$ and its harmonics.
The Blazhko modulation manifests itself as equidistantly spaced multiplets around the main frequency and its harmonics (Szeidl \& Jurcsik 2009; Benk\H{o} et al. 2009). All these frequencies were found through successive prewhitening, and the triplet components are strongly present in the data.
Therefore, in a first stage, we fit the data with a Fourier sum of the form:
\begin{equation}
\renewcommand{\arraystretch}{1.2}\begin{array}{l}
f(t) = A_0 + \sum_{k=1}^n [A_k \sin (2 \pi (kf_0 t + \phi_k))\\
+ A_k^{+} \sin (2 \pi ((kf_0 + f_{\rm B})t + \phi_k^{+}))\\
+ A_k^{-} \sin (2 \pi ((kf_0 - f_{\rm B})t + \phi_k^{-}))]\\
+ B_0 \sin (2 \pi (f_{\rm B} t + \phi_B)), \label{eq}
\end{array}
\end{equation}
From the combined Q1 and Q2 data, we find $f_0$=1.76416 $\pm$ 0.00001 d$^{\rm -1}$ for the main frequency, and its detected harmonics are significant up to the 14th order.
From the side peaks (triplet components) we find $f_{\rm B}$=0.0256 $\pm$ 0.0002 d$^{\rm -1}$ for the Blazhko frequency, or a Blazhko period of 39.1 $\pm$ 0.3 d.

From the Q2 data alone we find $f_0 = 1.76422 \pm 0.00001$ d$^{\rm -1}$, and $f_{\rm B} = 0.0252 \pm 0.0001$ d$^{\rm -1}$, or a Blazhko period of 39.6 $\pm$ 0.3 d.
As mentioned above, we also fit the Q2 data (Table~\ref{RR_freq}) with the values derived from Q1 and Q2 because of the longer time base.

\begin{figure}
\includegraphics[width=9cm]{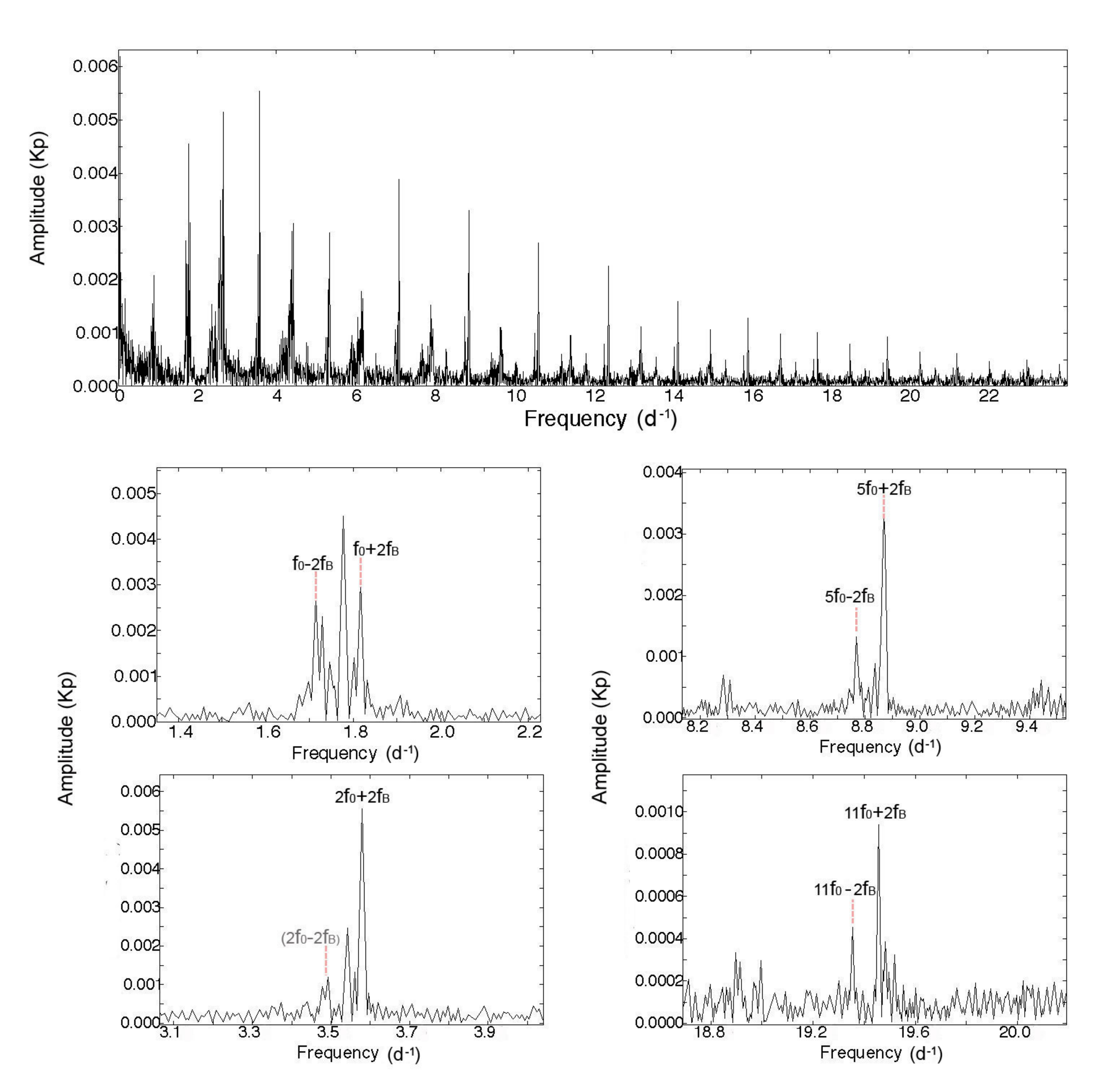}
 \caption{Residual spectrum of the {\it Kepler} RR Lyr data after prewhitening with the main frequency, its harmonics and the triplet components (top panel).  Zooms around some of the quintuplet components are shown in the bottom panels.\label{quint}}
\end{figure}

The main pulsation frequency we obtain is an {\it average} value of the actual, varying pulsation frequency over the time interval considered.  In fact, the amplitude {\it and} period of RR~Lyr's pulsation change significantly over the star's Blazhko cycle(s), as is shown in Fig.\,\ref{peramp}, obtained with the analytical signal method (Koll\'ath et al. 2002). The pulsation period varies between  0.5652\,d (1.7693 d$^{\rm -1}$) and 0.5699\,d (1.7547 d$^{\rm -1}$).  This corresponds to a period change $(\delta P/P$) of about 0.83 per cent, or nearly 12 minutes.
The period change is not identical from one Blazhko cycle to the next (see thick line in Fig.\,\ref{peramp} for consecutive Blazhko cycles).
The Blazhko frequency may in fact also be variable, and the value we obtain depends on the coverage (see Section\,\ref{Bl}).
Note that the harmonics of the Blazhko frequency ($2f_{\rm B}, 3f_{\rm B}$, etc.) do not reach the significance level in this data set, as they do in several other Blazhko stars (see, e.g., Chadid et al. 2010).  This may change with additional data of the star.

\subsection{Quintuplet components}
After subtracting a fit of the form of Equation\,\ref{eq}, we see many additional peaks in the spectrum. 
First of all, we see clear evidence for quintuplet frequencies. Fig.\,\ref{quint} shows the residual spectrum after prewhitening with the main frequency, its harmonics and the triplet components (top panel), and zooms around some of the quintuplet components (bottom panels).  The quintuplet components are added to Table\,\ref{RR_freq}. They become more easily discernible in the frequency spectrum with increasing order.  At low order, e.g., around $f_0$ and $2f_0$, other close frequencies have higher amplitudes (see Fig.\,\ref{quint}).
There are no clear indications for the occurrence of multiplet components of order higher than the quintuplet in the Fourier spectrum of the {\it Kepler} data of RR Lyr so far.

\begin{figure*}
\includegraphics[width=14cm]{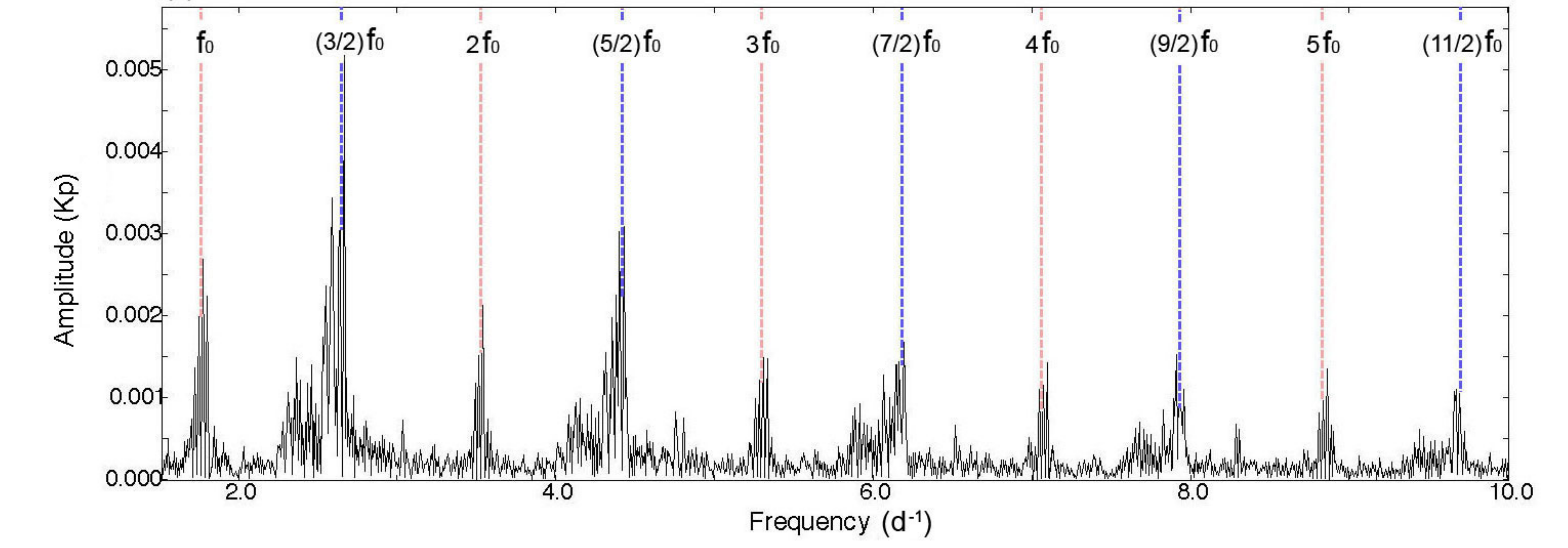}
 \caption{Fourier spectrum of the {\it Kepler} RR~Lyr data in the range [1.5 -- 10] d$^{\rm -1}$, after subtraction of the main frequency, its harmonics, and the triplet and quintuplets components. The positions of the mean pulsation frequency and its harmonics $kf_0$, and that of the half-integer frequencies $\frac{2k-1}{2}f_0$, are indicated by vertical lines. \label{RR_spec}}
\end{figure*}

\subsection{Period doubling}
It is not only around the main frequency and its harmonic components where we see additional peaks.
After prewhitening with the main frequency $f_0=1.76416$  d$^{\rm -1}$, its harmonics and the significant multiplet components, we clearly detect other significant frequencies in the Fourier spectrum. Fig.\,\ref{RR_spec} shows a zoom in the frequency range [1.5 -- 10] d$^{\rm -1}$ ($f_0$ -- $5f_0$). We clearly see additional signal around the half-integer multiples of $f_0$.  The occurrence of such half-integer components is connected to the period doubling phenomenon (see Section\,\ref{HIF}). As a consequence of period doubling we see the previously mentioned alternating higher and lower maxima, as can be seen in Fig.\,\ref{peramp} (most striking in the middle panel), in fig.\,4 of Kolenberg et al. (2010a) and in figs\, 1, 2 and 4 of Szab\'o et al. (2010). 

The highest peak is detected near 3/2$f_0$, followed by  5/2$f_0$ and  1/2$f_0$. The frequency at 2.6643 $\pm$ 0.0026 d$^{\rm -1}$  is the highest among the peaks around  3/2$f_0$. Its ratio to the main frequency is $1.510 \pm 0.001$, so there is a significant  deviation from the exact half-integer ratio.  We also observe several other frequency peaks around the 3/2$f_0$ component, and around the other half-integer frequency components, 
very close to each other, as shown in Fig.\,\ref{RR_spec}. They form asymmetric bunches that are more peaked towards the right (higher frequency) side.

\subsection{Additional peaks}

 After prewhitening with the main frequency, its harmonics, the observed triplet/quintuplet side peaks and the half-integer components, we observe many residual peaks around their positions (see Fig.\,\ref{RR_spec}).  Their spacing to the prewhitened frequency peaks is not equal or suspiciously close to the Blazhko frequency, so we cannot directly connect them with the Blazhko modulation.  A changing Blazhko effect, due to longer cycles, secular trends, or transient
phenomena would give rise to frequencies close to the typical pattern.  Further {\it Kepler} data to come will clarify what is the origin of the additional peaks.

\subsection{Light curve modulation}

\subsubsection{Modulation components}

\begin{figure}
\includegraphics[width=85mm, angle=0]{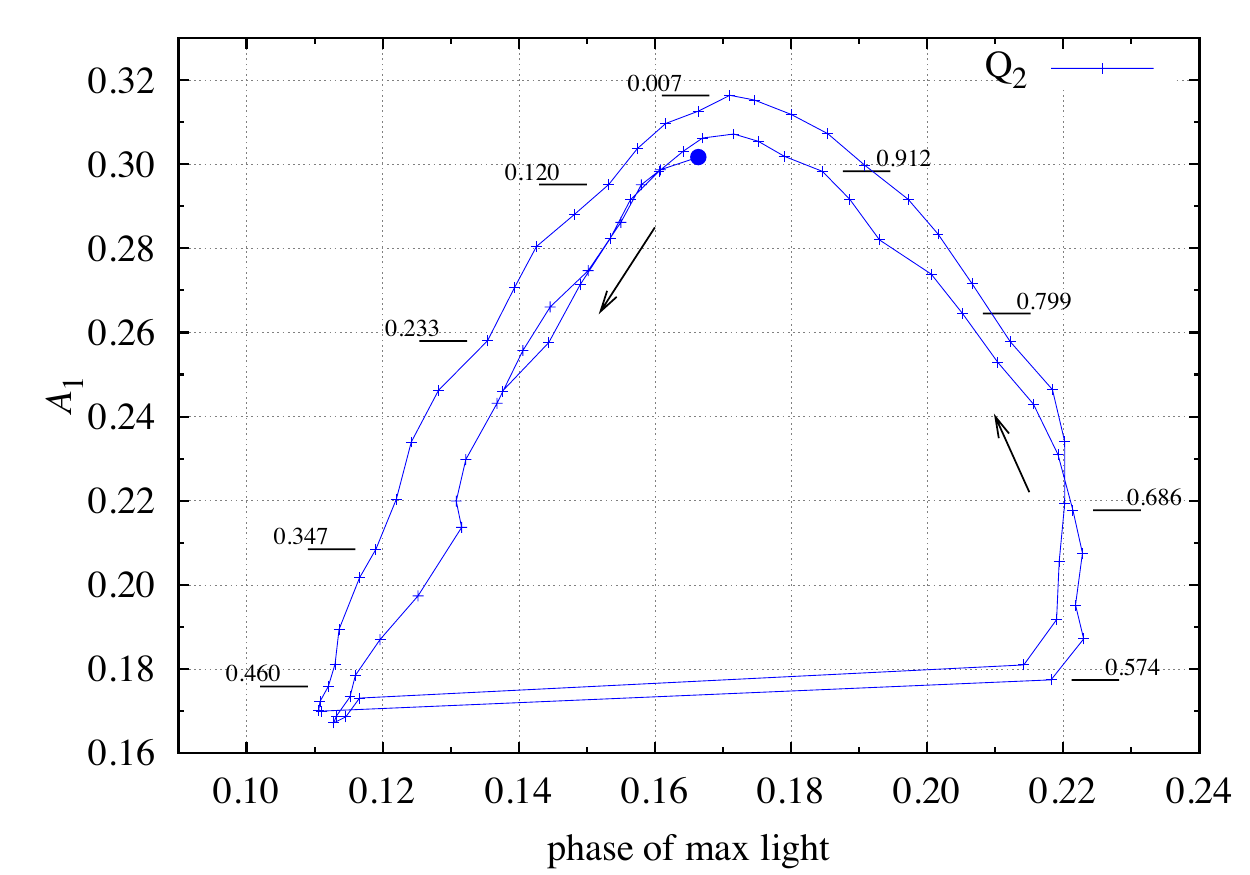}
 \caption{Amplitude variation versus phase of maximum light. The big dot marks the beginning of the trajectory. \label{phasefig}}
\end{figure}

The properties of the modulation components, specifically the triplet components $kf_0+f_{\rm B}$ and $kf_0-f_{\rm B}$, in the
frequency spectra of a Blazhko star can provide constraints for the models proposed to explain the modulation.  
Alcock et al. (2003) found that the relative amplitudes of the first order modulation components are usually in the range $0.1<A_1^{\pm}/A_1<0.3$.
From the {\it Kepler} data we find  $A_1^{+}/A_1 = 0.267 \pm 0.010$ and $A_1^{-}/A_1 = 0.049 \pm 0.007$.

\begin{figure*}
\includegraphics[width=160mm, angle=0]{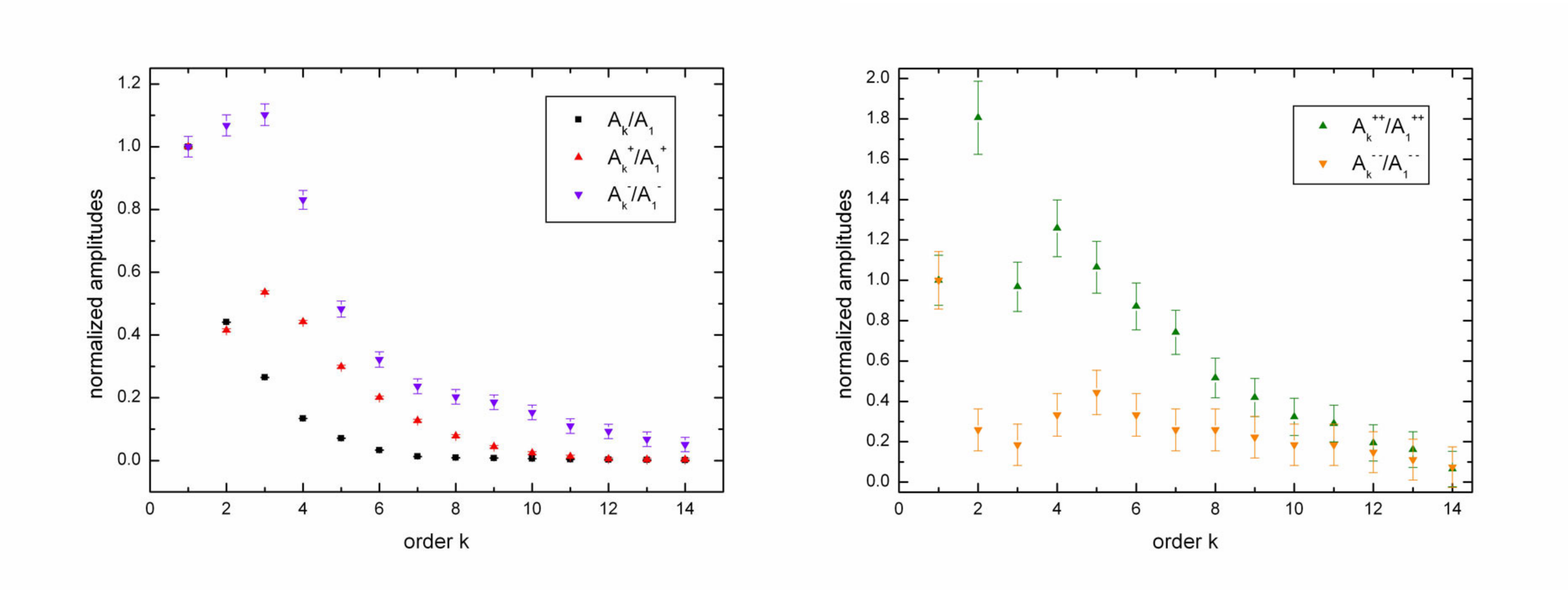}
 \caption{Left panel: amplitude ratios $A_k/A_1$,  $A_k^{+}/A_1^{+}$, and  $A_k^{-}/A_1^{-}$ of the detected harmonic and triplet frequencies in RR~Lyr, with error bars. Right panel: amplitude ratios $A_k^{++}/A_1^{++}$, and  $A_k^{--}/A_1^{--}$ of the detected quintuplet frequencies in RR~Lyr, with error bars.\label{ModCompDecrease}}
\end{figure*}

\begin{figure*}
\includegraphics[width=17cm, angle=0]{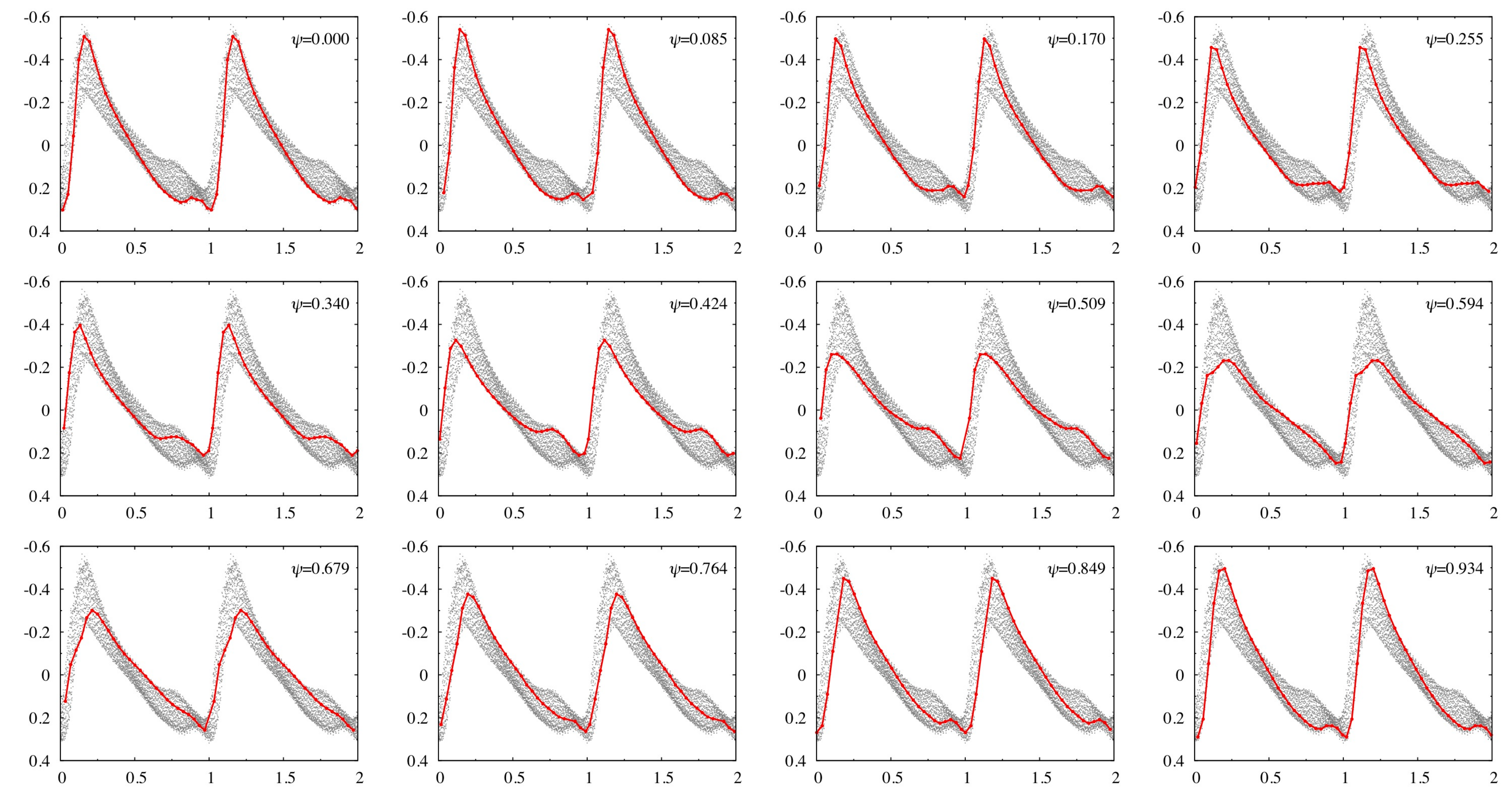}
 \caption{Variation of RR~Lyr's {\it Kepler} light curve (magnitude versus phase) for the second full Blazhko cycle in Q2, showing 12 snapshots (connected points) roughly equally spaced in time; each sixth pulsation cycle is plotted.  Blazhko phase $\psi = 0$ corresponds to maximum light amplitude. The dots are all the data in the Blazhko cycle folded with the main period. \label{LC}}
\end{figure*}

Table\,\ref{mod} lists the amplitude ratios, phase differences and their errors for
the first ten modulation component pairs, defined as
$R_k = A_{k}^{+}/A_{k}^{-}$ and $\Delta \phi_k = \phi_{k}^{+}-\phi_{k}^{-}$,
with $A_{k}^{\pm}$ and $\phi_{k}^{\pm}$ defined by Eq.\,1.
We also list the so-called asymmetry parameter $Q_k=\frac{A_k^{+}-A_k^{-}}{A_k^{+}+A_k^{-}}$,
introduced by Alcock et al. (2003) to quantify the degree of asymmetry in the peaks.
The distribution of the $Q$ parameter for the
Blazhko stars from the MACHO data base (Alcock et al. 2003) peaks at +0.3. 
The positive $Q$ points to an asymmetry with higher amplitudes at the higher
frequency lobes in the triplets, as is mostly the case in Blazhko stars.  
Finally, we list the power difference of the side peak amplitudes, $\Delta A_k^2 = (A_k^{+})^2-(A_k^{-})^2$. 
As Szeidl \& Jurcsik (2009) showed in their mathematical description of periodically modulated sinusoidal oscillation, this quantity depends on the phase difference between amplitude and phase modulation components.  Hence, it is the more physically meaningful quantity to measure the asymmetry of the triplet than the amplitude ratios $R_k$.
If the phase difference between the amplitude and phase modulations lies between 0 and $\pi$, then $A_1^{+} > A_1^{-}$ and the plot showing amplitude variation versus phase of maximum light has an anticlockwise progression, and vice versa (see Fig.\,\ref{phasefig}). 
Table\,\ref{mod} shows, through its values of $R_k > 1$ and $\Delta A_k^2$, that the triplet components in RR~Lyr always show asymmetry to the right side.
However, the asymmetry of the side peak amplitudes decreases towards higher orders, reflected in increasingly smaller $R_k$ and $\Delta A_k^2$ values.  Also their phase difference $\Delta \phi_k$, after peaking at order $k=2$, decreases towards higher orders.

\input{RR_tab2.tex}\label{mod}

Jurcsik et al. (2005) pointed out that generally the side peaks show a less steep decrease in amplitude towards higher orders than the
successive harmonics of $f_0$.  
Fig.\,\ref{ModCompDecrease} (left panel) shows the amplitude ratios $A_k/A_1$,  $A_k^{+}/A_1^{+}$, and  $A_k^{-}/A_1^{-}$. We confirm the less steep, though not ``linear" decrease for the modulation components in RR~Lyr.  The decrease at higher orders is less steep for the left sidepeak components. For the first time, we also show a similar figure for the quintuplet components (Fig,\,\ref{ModCompDecrease}, right panel).  Their decrease, especially at higher orders, is similar to that of the triplet components.  

\subsubsection{Fourier parameters}

RR~Lyr's light curve dramatically changes its shape over the Blazhko cycle. Fig.\,\ref{LC} shows the variation of RR~Lyr's light curve, through 12 snapshots evenly taken over the Blazhko cycle.  

We applied a time-dependent Fourier analysis (Kov\'acs et al. 1987) to derive the Fourier parameters for subsets of data. The data set was subdivided into segments consisting of either one pulsation cycle (26--28 points in a group), or four pulsation cycles (109--111 points in a group), where the pulsation cycle was defined starting from minimum brightness. The data were phased with a constant period, the average period $P = 0.566843$ d obtained from our frequency analysis.  A seventh-order harmonic fit, applied to these subsets of data, proved to yield the best fits for our sampling. The so derived Fourier parameters, i.e., the amplitude ratios $R_{k1} = A_{k}/A_{1}$  and the epoch-independent phase differences $\varphi_{k1} = \varphi_k-k\varphi_1$, offer a way to quantify the shape of the light curves.
We also used the analytical signal method (Koll\'ath et al. 2002) to analyze the data, yielding consistent results, but due to the gaps in the data we used the time-dependent Fourier analysis to present the continuous variation over the whole Blazhko cycle.

\begin{figure}
\includegraphics[width=85mm, angle=0]{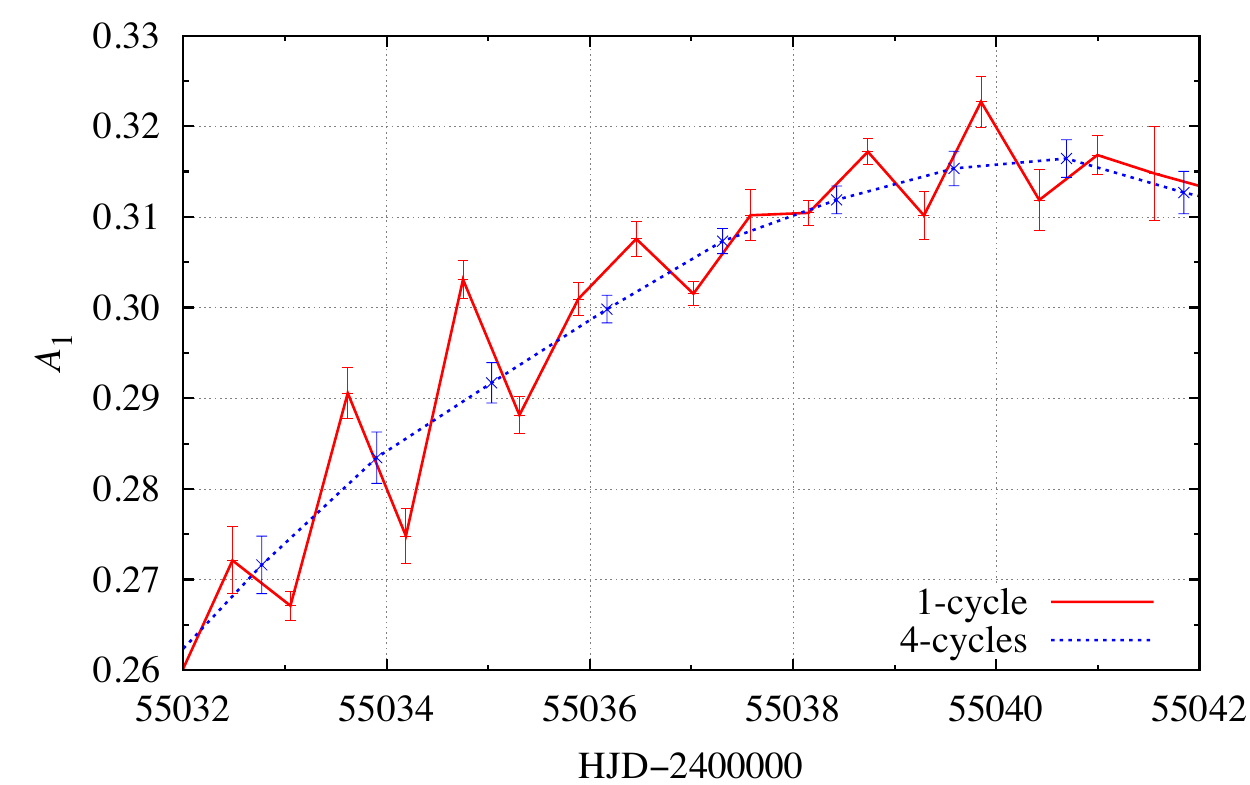}
 \caption{Comparison of the Fourier parameters (FP) $A_1$ derived from one-cycle and from four-cycle segments.  The effect of period doubling is clear in the parameters derived from one-cycle segments.  The oscillation of the parameter $A_1$ reflects the alternating height of successive light curves.  When taking four-cycle segments this effect is smoothed out. \label{PD}}
\end{figure}

Time-dependent Fourier parameters derived from segments four periods
long show a smooth time variation revealing the Blazhko time scale;
this procedure averages out the period-to-period oscillation 
caused by the period doubling phenomenon which is seen when segmenting
on single periods (Fig.\,\ref{PD}).

\begin{figure*}
\includegraphics[width=16cm, angle=0]{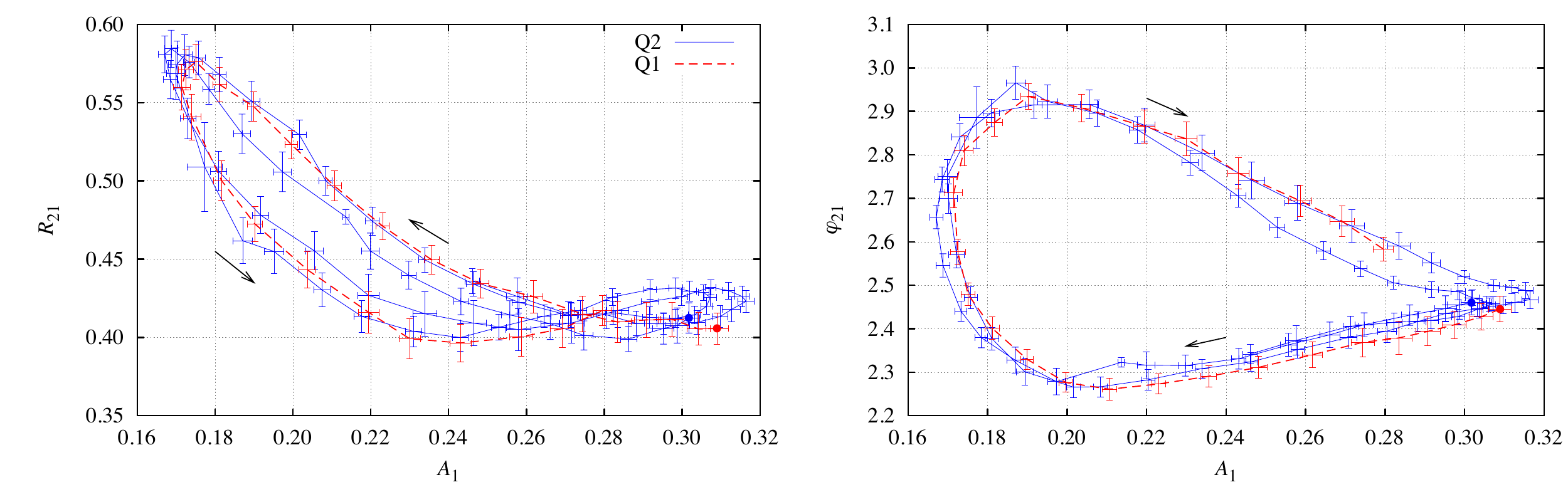}
 \caption{Interrelations between the Fourier parameters for successive Blazhko cycles, with error bars.  The Q1 data are connected with a dashed line, the Q2 data with a solid line. Arrows show the plot's progression. In addition, big
dots mark the beginning of trajectories. There is no exact repetition from one Blazhko cycle to the next. 
\label{FPfig}}
\end{figure*}

To show the light curve changes over the Blazhko cycle we use the four-cycle groups (Fig.\,\ref{FPfig}), thus averaging out the period doubling effect.  The interrelations between $A_1$ and $R_{21}$ (amplitude -- amplitude ratio) and $A_1$ and $\varphi_{21}$ (amplitude -- phase difference) describe loops with particular shapes, which models for the Blazhko effect should be able to reproduce. The $A_1$ -- $R_{21}$ interrelation shows a double loop, of which the largest part has a counter-clockwise progression.  For $A_1$ -- $\varphi_{21}$ the (single) loop is wider towards lower $A_1$ values, and its progression is clockwise. The loops for consecutive Blazhko cycles do not completely overlap: there is no exact repetition from one Blazhko cycle to the next.

\begin{figure}
\includegraphics[width=80mm, angle=0]{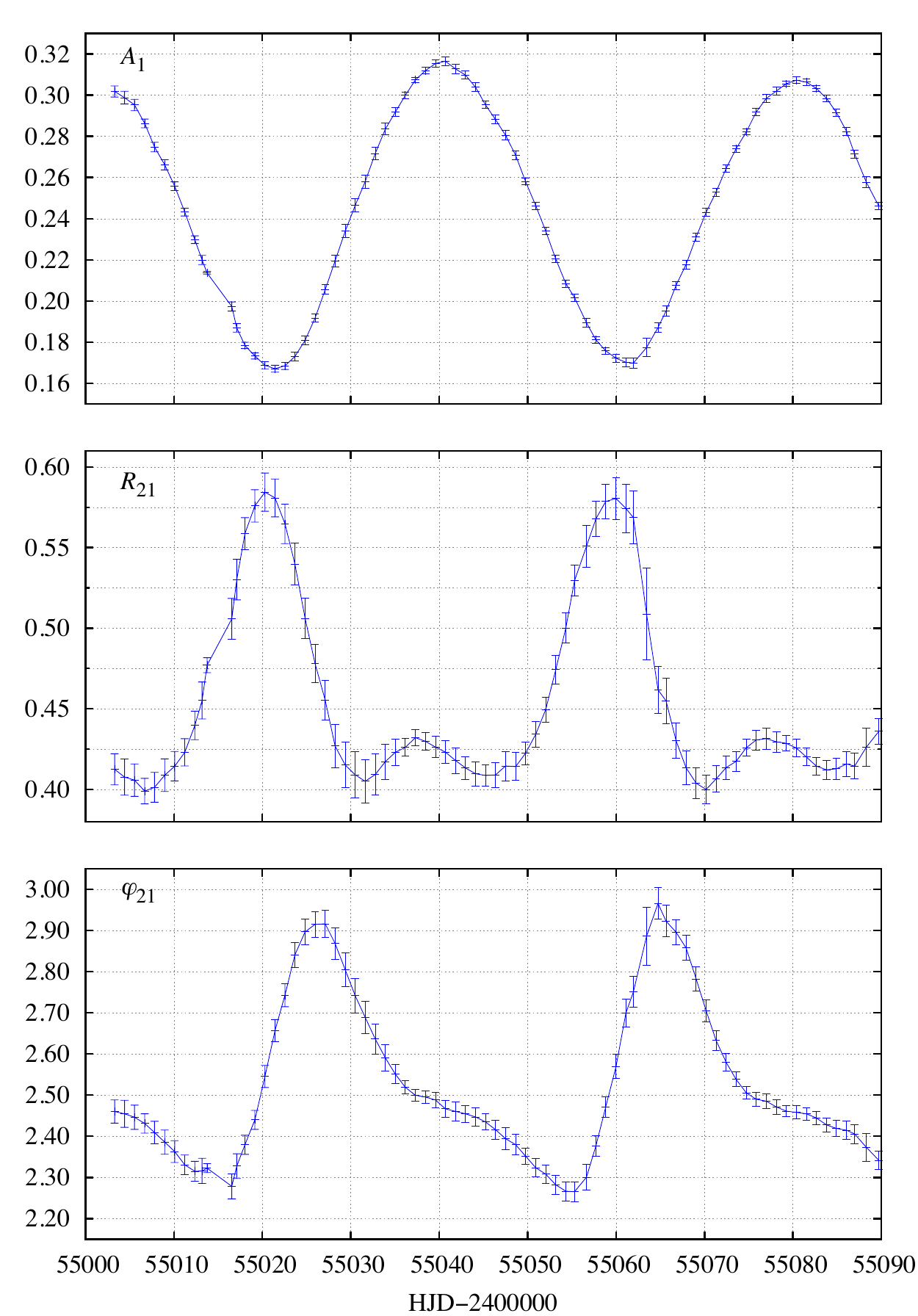}
 \caption{Variation of the Fourier parameters during the Blazhko modulation.\label{A1R21}}
\end{figure}

Fig.\,\ref{A1R21} shows the variation of the Fourier parameters during the modulation.
The variability of $A_1$ is almost sinusoidal, which 
shows that the modulation itself is not very nonlinear. 
$R_{21}$ reaches its maximum
when $A_1$ is at minimum. This is an observation that imposes constraints on the theoretical models for the Blazhko effect.
Generally, stars with smaller amplitudes have more sinusoidal
light curves. If during the modulation pulsation
energy is removed from the mode, the star is expected to behave as if being excited
to a lower amplitude: its variability should be more sinusoidal -- closer to the linear
regime -- and $R_{21}$ should decrease. We observe the opposite: the pulsation
is more nonlinear at smaller amplitudes. This may suggest that the modulation is caused by a mode resonance.
Detailed hydrodynamical models are needed to investigate the origin of this nonlinearity.

\section{Discussion}

\subsection{Blazhko modulation}\label{Bl}

For many decades, RR~Lyr's Blazhko period was known to be 40.8\,d (see, e.g., Szeidl \& Koll\'ath 2000). Kolenberg et al. (2006, their table\,6) list a summary of the Blazhko periods of RR~Lyr quoted in previous papers devoted to the star.  Unfortunately, error bars are rarely given and hard to reconstruct.  In the course of the past decades, the Blazhko period appears to have shortened. 
From the Q1 and Q2 {\it Kepler} data we find a value for the Blazhko period
$P_{\rm B}=39.1 \pm 0.3$ d, corresponding to a Blazhko frequency of $f_{\rm B}=0.0256$ d$^{\rm -1}$. 
From the Q2 {\it Kepler} data alone, we find $P_{\rm B}=39.6 \pm 0.3$ d, corresponding to a Blazhko frequency of $f_{\rm B}=0.0252$ d$^{\rm -1}$.

It is well established that periodic amplitude and phase modulation results in equidistantly spaced multiplets (Szeidl \& Jurcsik 2009; Benk\H{o} et al. 2009).
  Though RR~Lyr is one of the best studied stars of its class, it is the first time a significant detection of quintuplet components is reported in the star.  Since the first observation of quintuplet components by Hurta et al. (2008) in RV UMa, higher-order multiplet components (quintuplets and higher) have been found by several authors in high-quality data of Blazhko stars, such as MW~Lyr (Jurcsik et al. 2008), SS~For (Kolenberg et al. 2009) and the {\it CoRoT} target V1127 Aql (Chadid et al. 2010). 
  
As shown by Jurcsik et al. (2009), the mean parameters of a Blazhko RR Lyrae star vary over the course of the Blazhko cycle, such as the star's mean temperature, luminosity and radius.  Due to these changes, also the star's intrinsic period, determined from the $Kp$ light variations, varies between 0.5652\,d and 0.5699\,d (Fig.\,\ref{peramp}) for the Q2 data. As the Blazhko modulation appears to vary from one Blazhko cycle to the next, it is likely that the Blazhko period itself is also variable.

\subsection{Period doubling}\label{HIF}

With the {\it Kepler} photometry, the period doubling phenomenon in RR Lyrae stars was reported for the first time (Kolenberg et al. 2010a).
Even for a star as well-studied as RR~Lyr, it had never been detected before.  This is not only due to the much higher accuracy of the {\it Kepler} data, but also to the fact that from the mostly single-site observations of the star, consecutive pulsation cycles cannot be observed for stars with periods of typically half a day. The alternating higher and lower maxima would be easily missed by single-site observations due to daily gaps. Also recent multi-site observing campaigns of the star (e.g., Kolenberg et al. 2006) did not lead to a detection of the effect.

The period doubling and alternating heights of maxima are securely detected in two other {\it Kepler} stars, namely V808~Cyg (KIC4484128) and V355~Lyr (KIC755345) (see also Szab\'o et al. 2010).  A further four Blazhko RR Lyrae stars in the {\it Kepler} field possibly exhibit the effect: V2178 Cygni (KIC3864443), V354 Lyrae (KIC6183128), V445 Lyrae (KIC6186029) and V360 Lyrae (KIC9697825) (see Benk\H{o} et al. 2010).
So far the period doubling phenomenon has only been found in Blazhko stars and, though not all of our {\it Kepler} Blazhko stars seem to show it, there may be a connection between its occurrence and the Blazhko effect (Szab\'o et al. 2010).
The period doubling does not maintain the same strength along the Blazhko cycle. In some parts of the Blazhko cycle, the alternating light curve shapes are very obvious, in others indiscernible (see Fig.\,\ref{peramp}).

It is also of interest that period doubling occurs in RV~Tauri stars and in many
     hydrodynamical models of Pop.I (e.g., Moskalik \& Buchler 1991; Aikawa 2001) and Pop.II cepheids (e.g., Buchler \& Kovacs 1987; Buchler \& Moskalik 1992). 
 
 We refer to Szab\'o et al. (2010) for a more detailed study of the period doubling phenomenon from the {\it Kepler} data. They also investigate and simulate the deviation of the ratio from exact half-integer values. 
The latter is due to a combined effect of the Blazhko modulation and the temporal onset and disappearance of the half-integer frequencies.They suggest that a 9:2 resonance between the radial mode and the 9th-order radial overtone is responsible for the period doubling.

     Moskalik \& Buchler (1990) studied the period doubling found in Pop.I and Pop.II cepheid models in detail.
     Through the analysis of the limit cycle stability they showed that
     the 3:2 (in Pop.I) and 5:2 (in Pop.II) resonances are
     responsible. They demonstrated the coincidence of the period doubling
     and the resonance center, and provided a description of the mechanism in which the
     resonance destabilizes the limit cycle and causes the
     bifurcation. 
     They showed that every half integer resonance, in principle,
     is capable of causing period doubling.  
     The detailed modeling of the period doubling in Blazhko RR~Lyrae stars will be discussed in a forthcoming paper by Koll\'ath et al..
 
\subsection{Residual frequencies and additional cycles}

The {\it Kepler} data of RR~Lyr indicate that there is no strict repetition from one Blazhko cycle to the next, as is also reflected in Figs.\,5 and \ref{FPfig}.  The meticulous efforts to reconstruct the light variation of RR~Lyr described in Section\,\ref{phot} corroborate the reality of this variation.
In their recent high-quality {\it CoRoT} data of V1127~Aql, Chadid et al. (2010) found a small but significant shift of the maximum phase over five consecutive Blazhko cycles.   On a longer time scale, irregularities in RR~Lyr's Blazhko cycle were mentioned earlier by, e.g., Preston et al. (1965). Szeidl (1988) reported on weaker and stronger cycles in the star.  From the Blazhko cycles covered by the data presented in this paper we also see variation, even from one Blazhko cycle to the next.
Such a variation may be due to additional cycles in the star that are yet to be unraveled, or to non-periodic changes. The fact that we see residual signal around the position of the main frequency and its harmonics after subtracting the average frequency and the multiplet components is most likely due to these variations (periodic or transient).
Finally, the large corrections needed to adjust the flux measurements during some phases (Fig. 3), particularly in Q2, are probably not critical with regard to the detected additional variations.  As pointed out in Section\,\ref{phot}, these corrections do not introduce significant ``scatter" in the frequency results.  Hence we can safely claim that most of the ``scatter" can be attributed to variations in the Blazhko effect from cycle to cycle.

RR~Lyr itself is reported to have a longer cycle, of about four years, as first mentioned by Detre \& Szeidl (1973).  At the end of such a four-year cycle, the star's Blazhko effect is reported to weaken in strength, after which  a new cycle starts with increased strength and with a shift  (of about 0.25) in the Blazhko phase $\psi$. 

Kolenberg et al. (2006, 2008) list ephemerides for the Blazhko phase based on data gathered over the past decade.
However, with a changing Blazhko period and large gaps in the data, it is nearly impossible to determine whether there has been a phase shift in the Blazhko effect at a distinct moment in time.  If the abrupt phase shift is a real phenomenon, we hope to be able to observe it in the {\it Kepler} data to come.
Due to the period doubling it is not trivial to establish precise ephemerides for RR~Lyr, especially with long-cadence data.  Forthcoming data gathered in short cadence will allow us to do so.

\subsection{Light curve changes and bump behaviour}

Fig.\,\ref{LC} also shows the periodic behaviour of the so-called ``bump" in the light curve that appears just before minimum light. The bump occurs at an earlier phase around Blazhko minimum and later
around Blazhko maximum. Also its strength shows a dependence on
the Blazhko phase: it is most distinct during Blazhko minimum and becomes weaker around Blazhko maximum.  A similar behaviour of the bump was also observed by  Kolenberg et al. (2009) in the Blazhko star SS~For.  Guggenberger \& Kolenberg (2007) investigated it more quantitatively for both SS~For and RR~Lyr.

The bump has been explained by a collision between
layers in the deep atmosphere, leading to shock waves that, depending on their direction, are explained by the ``infall model" or the ``echo model" (Hill 1972; Gillet \& Crowe 1988). 
In their seminal work on spectroscopic data of RR~Lyr, Preston et al. (1965) concluded that a
displacement of the shock forming region occurs over the Blazhko cycle.  This induces an earlier or later occurrence of the bump in the light curves. Our observations suggest that this collision may be stronger and earlier when the star is at minimum Blazhko phase.  According to the findings by Jurcsik et al. (2009) for MW Lyr, this is also the phase when the star's mean radius is smaller (see their fig.\,14), which could explain the earlier occurrence of the collision. 
Although the radiative hydrodynamical models of Stothers (2006) are very crude, especially regarding the amplitudes of small features, the phase of the bump in his models also occurs as observed when the pulsation amplitude is low.  So it may be that this phasing is a consequence of an overall lower light amplitude, with the ``echo" arriving at the surface sooner because it is running through a more compact star.

Moreover, the so-called ``hump", the shoulder on the rising branch of the light curve, only occurs at certain phases in the Blazhko cycle.   In another early paper,  Struve (1948) reported striking hydrogen emission line profiles in RR~Lyr at particular, always ``rising Blazhko" phases (increasing light amplitude of the cycle light curves). 
This happens to be the Blazhko phase interval when we see a more pronounced hump on the rising branch. The sudden slope change of the rising branch may be connected with layers colliding (hence emission in the spectra) and thus slowing down the motions (hence slope change in the light curve). With long cadence data the rising branch is covered with too few data points (often only three).  Both the bump and hump characteristics can be investigated in detail with short-cadence data of RR~Lyr.

\subsection{Constraints on the existing Blazhko models}\label{constraints}

How do the {\it Kepler} observations of RR~Lyr provide constraints for improving the existing Blazhko models?
The most striking features we found, for the first time, in this data set are the following:
\begin{enumerate}
\item We found additional frequencies: quintuplet components and, most importantly, the period doubling phenomenon.
\item We detected extreme light curve variations measured with high accuracy, e.g., the hump and bump feature. 
\item We observed that there is no strict repetition from one Blazhko cycle to the next, and that the Blazhko period may also vary on this short time scale.
\end{enumerate}
In addition, the most important findings so far from long cadence data of all 29 {\it Kepler} RR Lyrae stars are (Benk\H{o} et al. 2010):
\begin{enumerate}
\item About half of the stars, 14 out of 29 studies RR Lyrae stars in our {\it Kepler} sample, are modulated.
\item Period doubling and additionally excited higher-order radial overtones detected in seven stars, all of them Blazhko stars.
\item All 14 Blazhko stars show both period modulation {\it and} amplitude modulation, in differing degrees.
\end{enumerate}

With increasingly impressive data sets and subsequent new and detailed findings, it is clear that the models so far proposed for the Blazhko effect are lagging behind.
Any model for the Blazhko effect has to take into account both amplitude and phase modulation, as pointed out by Benk\H{o} et al. (2010). Moreover, explanations that strictly impose a single modulation period, (e.g., the star's rotation period) not allowing for variation of the modulation, are definitely no longer viable. The moving light curve features (hump, bump) can provide additional clues about the stellar dynamics varying over the Blazhko cycle.  Moreover, any plausible model has to be able to account for the (transient) occurrence of period doubling and possible excitation of higher order overtones.  

The recent findings by Szab\'o et al. (2010) in connection with the period doubling suggest a 9:2 resonance between the radial mode and the 9th-order radial overtone (a strange mode) to be responsible for the period doubling.   

In the scenario proposed by Stothers (2006) a variation of the star's convection is the cause for the Blazhko modulation.  As the star's mean stellar parameters (temperature, radius and luminosity) vary over the Blazhko cycle (see Jurcsik et al. 2009), one can reasonably assume that its turbulent/convective structure also changes. It is unclear, however, what could lie behind this change. 
We note that transient magnetic fields at the origin of the variation in convective turbulence, as postulated by Stothers (2006), are most probably undetectable with present-day instrumentation, and thus their existence would be hard to prove.

It is clear that it is time to revise or expand the existing models for the Blazhko effect, as well as explore alternative explanations, using the constraints described above.  Perhaps it is also time to revisit previously discarded scenarios, such as models based on {\it radial} mode resonances (see, e.g., Goranskij et al. 2009; Kov\'acs 2009), or explore radial hydrodynamical models in much more detail. After the models by Borkowski (1980) and 
Moskalik (1986) were dismissed by lack of confirmation from hydrodynamical models, 
most attention has been on resonances involving nonradial modes. 

\section{Conclusions and outlook}
The first releases of the {\it Kepler} data of the prototypical star RR Lyr are revolutionary because:
\begin{itemize}

\item These data prove that bright stars such as RR~Lyr can be observed with {\it Kepler}, as the saturated flux is conserved to a high degree (less
than $2.5 \times 10^{-3}$ fractional error even at maximum flux). 
Generally the philosophy is not to waste any pixels on a star of which one cannot capture all the useful flux.  
RR Lyr is an unusual case due to its high variability. Because there are measurements in which the star is well captured and there is a lot of saturated flux in columns captured in every measurement, we can make good estimates of missing flux and thus recover photometry.  Such unusual cases were not considered earlier in the {\it Kepler} Mission.

\item  From the three Blazhko cycles covered by the data set (Q1+Q2 data), we
  obtain $f_0 = 1.76416 \pm 0.00001 $ 
d$^{\rm -1}$ for the {\it average} radial pulsation frequency and 
$f_{\rm B} = 0.0256 \pm 0.0002$ 
d$^{\rm -1}$ for the 
Blazhko frequency. 
The Blazhko period is $P_{\rm B}=39.1 \pm 0.3$ d, which confirms the earlier findings that the star's modulation period has gradually become shorter over the past decades (Kolenberg et al. 2006).  

\item There are, however, also indications that the Blazhko period is variable on shorter time scales, because our data clearly indicate that there is no strict repetition from one Blazhko cycle to the next. 

\item We clearly detect quintuplet components in the frequency spectrum of the {\it Kepler} RR~Lyr data.  It is the first time these are found in the star.   Their behaviour is similar to that of the triplet components.

\item Moreover, we find clear evidence for additional frequencies in the data.  The most striking new frequency pattern is given by the so-called half-integer frequencies, that occur at $\frac{2k-1}{2}f_0$, with the highest peaks around $3/2f_0$, $5/2f_0$, $1/2f_0$, etc. Their occurrence is connected to the observation of alternating higher and lower maxima throughout (certain phase intervals of) the Blazhko cycle.  This phenomenon, called period doubling, has been observed for the first time in {\it Kepler} data (Kolenberg et al. 2010a).  Meanwhile, we have observed it in several Blazhko stars observed with {\it Kepler}, and its presence may be connected to the Blazhko effect. 
The period doubling is variable over the course of the Blazhko cycle, and also does not repeat identically from one Blazhko cycle to the next (Szab\'o et al. 2010). 

\item The Fourier parameters derived from individual cycles oscillate around the Fourier parameters derived from several consecutive cycles, for which the period doubling effect is smoothed out.  

\item   The increase of the Fourier parameter $R_{21}$ with decreasing pulsation amplitude $A_1$
is counterintuitive, because pulsations with lower amplitudes (lower $A_1$) are expected to be less nonlinear (lower $R_{21}$).  This observation provides constraints for modelling the physics of the Blazhko modulation.

\item The position and strength of the bump and hump vary over the Blazhko cycle. A detailed
  study of these features in the pulsation cycle, in photometric as well
  as spectroscopic data, may shed new light upon the pulsation and shock propagation in RR Lyrae stars and the understanding of the Blazhko effect.
  \end{itemize} 

After the Q2 observations, it was realized that RR~Lyr's {\it Kepler} KIC magnitude was not bright enough to match the observations, and it was adjusted upwards. As a consequence, the standard aperture assignment algorithms assigned it to an aperture that went off silicon. Thus the target was rejected, and was not observed in Q3 and Q4.  The first {\it Kepler} findings described in Kolenberg et al. (2010a), however, triggered further interest in the star.  Therefore, a custom aperture was devised for RR~Lyr (by Steve Bryson), reducing the amount of originally assigned pixels from 433 to about 150.  Thanks to the calibration work described in this paper, RR~Lyr will likely be scheduled to be a target of {\it Kepler} for the rest of the mission. At the time of writing of this paper, RR~Lyr is being observed through the {\it Kepler} Guest Observer (GO) program {\rm (http://keplergo.arc.nasa.gov/)}. 

\vspace{2mm}
As further {\it Kepler} data of RR~Lyr become available, we will be able to perform more focused analyses:

 \begin{itemize}
 
 \item A longer time base can lead to a better resolved frequency spectrum.  However, due to the period variation and the transient phenomena, the frequency spectrum will be smeared out. Hence, in reality we will probably not obtain ``sharper" peaks with a longer time base, but several closely-spaced peaks.  New and additional {\it Kepler} data will enable us to analyse the temporal variation of the frequencies. They will also help uncover whether the variation between consecutive Blazhko cycles is due to additional cycles in the star, or to transient or quasi-periodic changes.
  \item {\it Kepler} short-cadence data will allow us to study short-time effects and light curve features such as the hump and bump. Simultaneous ground-based data are planned for the period of the short-cadence {\it Kepler} observations.
 \item With {\it Kepler}'s potential coverage of RR~Lyr's light variations over 3.5--5\,y, many mysteries still surrounding this prototypical star can be solved.  We may witness the start of a new 4-year cycle (Detre \& Szeidl 1973), as RR~Lyr is known to have.
 \item With {\it Kepler} we can anticipate a dramatic overhaul in the models for the Blazhko effect.
 The constraints provided by the ultra-precise data of RR~Lyr motivate us to revisit, revise or expand the existing models for the Blazhko effect, as well as explore alternative explanations. 

\end{itemize}

\section*{Acknowledgments}

The authors kindly thank the anonymous referee for constructive comments.
Funding for this Discovery mission is provided by NASA's Science Mission Directorate. The authors gratefully acknowledge the entire {\it Kepler} team, whose outstanding efforts have made these results possible. 

KK and EG acknowledge  support from the Austrian Fonds zur F\"orderung der wissenschaftlichen Forschung, project number  T359-N16 and P19962.
RSz, JB and LLK are supported by the National Office for Research and
Technology through the Hungarian Space Office Grant No. URK09350, the
Lend\"ulet program of the Hungarian Academy of Sciences, and OTKA Grants K76816 and MB08C 81013. 
RS is supported by FWF project AP2120521.

\end{document}

%% file: RR_tab1.tex
\begin{table*}
\caption{Amplitudes and phases 
of the pulsation and modulation 
frequency components of RR~Lyr for the best fit to the Q2 data.  The values displayed in {\it italics}
correspond to combination frequencies not exceeding a signal-to-noise level 
of 3.5.  The residuals of the fit to the data are 0.0127 mag.  We also list the most prominent additional period doubling components that are found in the data set. 
The uncertainty on $f_0$ is  $1 \times 10^{-5}$ d$^{\rm -1}$ and the
uncertainty on $f_B$ is $2 \times 10^{-4}$ d$^{\rm -1}$. \label{RR_freq}}
\centering                          

\begin{tabular}{|lrllcl|lllrl|}        
\hline               
\multicolumn{2}{c}{$f$ [d$^{\rm -1}$]} &
\multicolumn{1}{l}{$A_{Kp}$ [mag]} &
\multicolumn{1}{l}{$\phi$ [$\frac{\rm rad}{2\pi}$]} &
\multicolumn{1}{l}{$\sigma(\phi)$}&
\multicolumn{1}{l}{}&
 \multicolumn{2}{|c}{$f$ [d$^{\rm -1}$]} &
\multicolumn{1}{l}{$A_{Kp}$ [mag]} &
\multicolumn{1}{l}{$\phi$ [$\frac{\rm rad}{2\pi}$]} &
\multicolumn{1}{l|}{$\sigma(\phi)$}
 \\
&& $\pm 0.0006$ & & & & && $\pm 0.0006$ & &  \\
\hline                     
$f_0$   &      1.76416 &     0.2392  &   0.1540 & 0.0003 & & &&&&\\
$2f_0$  &       3.52831 &     0.1056  &   0.7130 & 0.0007 & & & \multicolumn{4}{l}{Quintuplet components} \\
$3f_0$  &        5.29247 &    0.0633  &    0.3074 &0.0012 & & & &&& \\
$4f_0$  &      7.05663  &    0.0321 &    0.9124 & 0.0023 & &$f_0-2f_B$  &        1.71296 &   0.0027 &   0.9146 & 0.0263 \\
$5f_0$  &       8.82078  &   0.0170 &     0.5260 & 0.0040 &&$f_0+2f_B$  &        1.81535 &   0.0031 &   0.9481 & 0.0228 \\
$6f_0$  &       10.58494  &   0.0079 &    0.1885  & 0.0094 && $2f_0-2f_B$  &        3.47712 &   {\it 0.0007} &  {\it 0.4657} & {\it 0.0931} \\
$7f_0$  &       12.34910  &   0.0031 &   0.8782 & 0.0224 && $2f_0+2f_B$  &        3.57950 &   0.0056 &   0.0753 & 0.0111 \\
$8f_0$  &       14.11325  &   0.0021 &   0.5445  & 0.0327 &&$3f_0-2f_B$  &        5.24128 &   {\it 0.0005} &  {\it 0.5644} & {\it 0.1477} \\
$9f_0$  &       15.87741  &  0.0018 &    0.1971 &  0.0445 && $3f_0+2f_B$  &        5.34366 &   0.0030 &   0.6715 & 0.0228 \\
$10f_0$ &       17.64157  &  0.0015 &     0.8210 & 0.0533 && $4f_0-2f_B$  &        {\it 7.00544} &   {\it 0.0009} &  {\it 0.2866} & {\it 0.0724} \\
$11f_0$ &     19.40573  & 0.0009 &   0.4642 & 0.0756 && $4f_0+2f_B$  &        7.10782 &   0.0039 &   0.1162 & 0.0169 \\
$12f_0$ &      21.16988  & 0.0006 & 0.0843 & 0.1392 && $5f_0-2f_B$  &        8.76859 &   0.0012 &   0.9664 & 0.0611 \\
$13f_0$ &     22.93404  & 0.0004 &  0.6760 & 0.2313 && $5f_0+2f_B$  &        8.87198 &   0.0033 &   0.6446 & 0.0207 \\
{\it $14f_0$} &     {\it  24.69820} & {\it 0.0003} &   {\it 0.2974}   & {\it 0.2793} && $6f_0-2f_B$  &        10.53375 &   0.0009 &   0.5928 & 0.0846 \\
&&&& &&  $6f_0+2f_B$  &        10.63613  &   0.0027 &   0.1941 & 0.0255 \\
& \multicolumn{4}{l}{Blazhko frequency}&& $7f_0-2f_B$  &        12.29791 &   0.0007 &   0.1686 & 0.0869 \\
&&&& && $7f_0+2f_B$  &       12.40029 &   0.0023 &   0.7500 & 0.0287 \\
$f_B$   &  0.02560  & 0.0119 & 0.3220 & 0.0055 &&$8f_0-2f_B$  &        14.06206 &   0.0007 &   0.7379 & 0.1092 \\
&&&& && $8f_0+2f_B$  &        14.16445  &   0.0016 &   0.3262 & 0.0405 \\
& \multicolumn{4}{l}{Triplet components}&&$9f_0-2f_B$  &        15.82622 &   0.0006 &   0.3213 & 0.1264 \\
&&&& && $9f_0+2f_B$  &        15.92860  &   0.0013  &   0.8751 & 0.0516 \\ 
$f_0-f_B$  &       1.73856 &    0.0118  & 0.8730     &0.0054 && $10f_0-2f_B$  &        17.59038  &   0.0005 &   0.8891 & 0.1388 \\
$f_0+f_B$  &       1.78975 &   0.0638 &   0.2861     &0.0012 && $10f_0+2f_B$  &       17.69276 &   0.0010 &   0.4580 & 0.0649 \\
$2f_0-f_B$  &      3.50272 &      0.0126 &  0.2502 & 0.0055 && $11f_0-2f_B$  &        19.35453 &   0.0005 &   0.5111 & 0.1886 \\
$2f_0+f_B$  &      3.55391 &    0.0265 &   0.7785 &0.0027  && $11f_0+2f_B$  &        19.45691 &   0.0009 &   0.0300 & 0.0753 \\
$3f_0-f_B$  &      5.26687 &     0.0130  &  0.8662  & 0.0051&& $12f_0-2f_B$  &        21.11869 &   0.0004 &   0.1098 & 0.1939 \\
$3f_0+f_B$  &      5.31807 &  0.0342  &  0.3142  & 0.0020 && $12f_0+2f_B$  &        21.22107 &   0.0006 &   0.6290 & 0.1410  \\
$4f_0-f_B$  &     7.03103 &     0.0098 &   0.5406  &  0.0064&& $13f_0-2f_B$  &        {\it 22.88285} &   {\it 0.0003} &  {\it 0.7488} & {\it 0.2500} \\
$4f_0+f_B$  &     7.08222 &    0.0282   &  0.8995  &0.0025 && $13f_0+2f_B$  &       22.98523 &   0.0005  &  0.1849 & 0.1714 \\
$5f_0-f_B$  &      8.79519 &  0.0057 & 0.1974 &  0.0115 && $14f_0-2f_B$  &        {\it 24.64700} &  {\it 0.0002} &  {\it 0.3263} & {\it 0.3335} \\
$5f_0+f_B$  &     8.84638 &     0.0191  &  0.4783  &0.0036 && $14f_0+2f_B$  &       {\it 24.74938} &  {\it 0.0002} &  {\it 0.6737} & {\it 0.2951} \\
$6f_0-f_B$  &        10.55935 &  0.0038 & 0.8244 &  0.0180 && &&&&\\
$6f_0+f_B$  &       10.61054 &     0.0128  &  0.0550 &0.0054 && & \multicolumn{4}{l}{Most prominent half-integer frequencies} \\
$7f_0-f_B$  &       12.32350 &  0.0028 &    0.4399 &  0.0253 && &&&&\\
$7f_0+f_B$  &       12.37470 &     0.0081  &  0.6358 &0.0084 && $\simeq 1/2f_0$ & 0.89945 &   0.0019 & 0.7063 &   0.0320\\
$8f_0-f_B$  &        14.08766 &    0.0024 & 0.0508 & 0.0282 && $\simeq 3/2f_0$ & 2.66431 &  0.0052 & 0.1313 & 0.0122 \\
$8f_0+f_B$  &       14.13885 &     0.0050  & 0.2285 & 0.0143&& $\simeq 5/2f_0$ & 4.40118 & 0.0030 & 0.8108 & 0.0223 \\
$9f_0-f_B$  &       15.85182 &    0.0022 &  0.6648 & 0.324 && $\simeq 7/2f_0$ & 6.14280 & 0.0017 & 0.1515 & 0.0388 \\
$9f_0+f_B$  &     15.90301 &     0.0028  & 0.8169 & 0.224 && $\simeq 9/2f_0$ & 7.90739 & 0.0014 & 0.4453 & 0.0427 \\
$10f_0-f_B$  &       17.61597 &    0.0018 &   0.2852 & 0.0383  && $\simeq 11/2f_0$ & 9.66602 & 0.0010 & 0.5434 & 0.0650 \\ 
$10f_0+f_B$  &      17.66716 &     0.0015 &  0.4175 & 0.0483 && $\simeq 13/2f_0$ & 11.42924 & 0.0009 & 0.7997 & 0.0783 \\
$11f_0-f_B$  &        19.38013 &   0.0013 &   0.8873 & 0.0533 && $\simeq 15/2f_0$ & 13.22292 & 0.0011 & 0.6350 & 0.0668 \\
$11f_0+f_B$  &       19.43132 &   0.0008  & 0.0091 & 0.0804 &&   $\simeq 17/2f_0$ & 14.98651 & 0.0011 & 0.8630 & 0.0611 \\
$12f_0-f_B$  &        21.14429 &   0.0011 &   0.4774 & 0.0661&& $\simeq 19/2f_0$ & 16.75016 & 0.0009 & 0.7943 & 0.0708 \\
{\it $12f_0+f_B$}  &      {\it 21.19547} &    {\it 0.0003}  & {\it 0.6314} & {\it 0.3044}  && $\simeq 21/2f_0$ & 18.51361 & 0.0008 & 0.3654 & 0.0944 \\
$13f_0-f_B$  &       22.90844 &   0.0008 &  0.0825 &0.0839 && $\simeq 23/2f_0$ & 20.27817  & 0.0006 & 0.8340 & 0.1100 \\
{\it $13f_0+f_B$}  &      {\it 22.95964} &  {\it  0.0002} & {\it 0.2807} & {\it 0.3710} && $\simeq 25/2f_0$ & 22.04152 & 0.0004 & 0.9943 & 0.1847 \\
$14f_0-f_B$  &     24.67260 &   0.0006 &  0.6800 & 0.1304 && $\simeq 27/2f_0$ & 23.80626 & 0.0004 & 0.5284 & 0.1915 \\
{\it $14f_0+f_B$}  &      {\it 24.72379} &    {\it 0.0002}  & {\it 0.2037} & {\it 0.3488} && &&&&\\

\hline

\end{tabular}
\end{table*}

%% file: RR_tab2.tex
\begin{table*}
\caption{Side lobe amplitude ratios $R_k$, phase differences $\Delta \phi_k$
($\times 2\pi$), asymmetry
  parameters $Q_k$, and power differences  $\Delta  A_k^2 $ as defined in the
  text, and their respective
  errors (based on Monte Carlo errors) for the $Kp$ data of RR~Lyr.  
$k$ denotes the multiplet order. Values with a significance below $3\sigma$ are given in {\it italics}.  
\label{mod}  }    
\centering                          

\begin{tabular}{c c c c c c c r r}        
\hline               
$k$ & $R_k$ & $\sigma_{R_k}$ & $\Delta \phi_k $ & $\sigma_{\Delta \phi_k}$ &
$Q_k$ & $\sigma_{Q_k}$  & $\Delta  A_k^2 $ &  $\sigma_{\Delta  A_k^2 }$\\
\hline      
1	&	5.41	&	0.28	&	2.60	&	0.03	&	0.69	&	0.01	&	0.0039312	&	0.0000551	\\
2	&	2.10	&	0.11	&	3.32	&	0.04	&	0.36	&	0.02	&	0.0005435	&	0.0000249	\\
3	&	2.63	&	0.13	&	2.81	&	0.03	&	0.45	&	0.02	&	0.0010006	&	0.0000310	\\
4	&	2.88	&	0.19	&	2.26	&	0.04	&	0.48	&	0.02	&	0.0006992	&	0.0000253	\\
5	&	3.35	&	0.37	&	1.76	&	0.08	&	0.54	&	0.04	&	0.0003323	&	0.0000169	\\
6	&	3.37	&	0.55	&	1.45	&	0.12	&	0.54	&	0.06	&	0.0001494	&	0.0000113	\\
7	&	2.89	&	0.66	&	1.23	&	0.17	&	0.49	&	0.09	&	0.0000578	&	0.0000073	\\
8	&	2.08	&	0.58	&	1.12	&	0.20	&	{\it 0.35}	&	{\it 0.12}	&	0.0000192	&	0.0000047	\\
\hline
\end{tabular}
\end{table*}